\documentclass{optica-article}

\journal{opticajournal} 

\articletype{Research Article}
\usepackage{amsmath}
\usepackage{physics}
\usepackage{lineno}
\usepackage{hyperref}

\begin{document}

\title{High-dimensional quantum key distribution using a multi-plane light converter}

\author{Ohad Lib\authormark{1,$\dag$,*}, Kfir Sulimany\authormark{1,$\dag$},Mateus Araújo\authormark{2},Michael Ben-Or\authormark{3}, and Yaron Bromberg\authormark{1}}

\address{\authormark{1}Racah Institute of Physics, The Hebrew University of Jerusalem, Jerusalem 91904, Israel}
\address{\authormark{2}Departamento de Física Teórica, Atómica y Óptica, Universidad de Valladolid, 47011 Valladolid, Spain}
\address{\authormark{3}School of Computer Science \& Engineering, The Hebrew University of Jerusalem, Jerusalem, 91904 Israel}
\authormark{$\dag$}The authors contributed equally to this work.
\email{\authormark{*}ohad.lib@mail.huji.ac.il} 

\begin{abstract*} 
High-dimensional quantum key distribution (QKD) offers higher information capacity and stronger resilience to noise compared to its binary counterpart. However, these advantages are often hindered by the difficulty of realizing the required high-dimensional measurements and transformations. Here, we implement a large-scale multi-plane light converter (MPLC) and program it as a high-dimensional mode sorter of spatial modes for QKD. Using the MPLC, we demonstrate five-dimensional QKD with six mutually unbiased bases and 25-dimensional QKD with two mutually unbiased bases in the same experimental setup. Furthermore, we propose a construction of pairs of mutually unbiased bases that are robust to experimental errors, with measurement complexity scaling only with the square root of the encoded dimension. This approach paves the way for QKD implementations in higher dimensions.
\end{abstract*}

\section{Introduction}

Quantum key distribution (QKD) facilitates the sharing of a secret key between two parties, Alice and Bob, by encoding and sending quantum information between them\cite{bennett1984quantum,ekert1991quantum}. It is typically implemented by encoding quantum bits (qubits) onto the polarization or time-of-arrival of photons and was successfully demonstrated in a variety of experiments, including through long fiber-optic\cite{wang2022twin} or free-space\cite{schmitt2007experimental,liao2017satellite} links. Unfortunately, loss and noise in the optical link substantially reduce the achievable secure key rates in such systems\cite{xu2020secure}, hindering their widespread use. 

A promising approach for increasing the information capacity of QKD systems and improving their resilience to noise is through high-dimensional encoding\cite{cerf2002security,erhard2018twisted,erhard2020advances}. In these systems, high-dimensional degrees of freedom are used to encode d-dimensional quantum states (qudits) that are then measured in two or more mutually unbiased bases (MUBs). High-dimensional quantum key distribution protocols have been demonstrated utilizing spatial \cite{walborn2006quantum,etcheverry2013quantum,mirhosseini2015high,ding2017high,sit2017high,bouchard2017high,bouchard2018experimental,cozzolino2019orbital,tentrup2019large,zhou2019using,otte2020high,da2021path,hu2021pathways,ortega2021experimental,stasiuk2023high,halevi2024high}, time-bin\cite{islam2017provably,lee2019large,vagniluca2020efficient,ikuta2022scalable,chapman2022hyperentangled,sulimany2021fast}, or time-energy \cite{ali2007large,mower2013high,lee2014entanglement, zhong2015photon,liu2019energy,bouchard2021achieving,liu2023high,bulla2023nonlocal,chang2023large} encoding. Spatial encoding is particularly attractive for high-dimensional QKD, as spatial light modulators (SLMs) allow efficient control over such high-dimensional states, which can then be transmitted through either multimode fibers\cite{cozzolino2019orbital,cao2020distribution,sulimany2022all}, multi-core fibers\cite{da2021path,ortega2021experimental}, or free-space links\cite{vallone2014free,sit2017high}.

However, the experimental complexity associated with measuring such spatially encoded qudits in multiple MUBs limits the performance of high-dimensional QKD. These measurements are often performed probabilistically: the state received by Bob is projected onto one out of $d$ possible states, making the measurement success rate inversely proportional to the dimension $d$. A better approach utilizes mode sorters, which direct each state to a different detector, ultimately enabling perfect measurements regardless of the dimension\cite{mirhosseini2015high}. While mode sorters have been recently reported, realizing programmable mode sorters of spatial modes in multiple high-dimensional MUBs remains a major challenge.   

Recently, multi-plane light converters (MPLCs) have emerged as a promising tool for realizing large, programmable mode sorters, using a cascade of optimized phase masks separated by free-space propagation\cite{morizur2010programmable}. The large number of programmable phase pixels in SLMs, together with the good connectivity provided by diffraction, make MPLCs a flexible platform that suits various applications\cite{morizur2010programmable,labroille2014efficient,kupianskyi2023high,kupianskyi2024all}, ranging from classical communication\cite{fontaine2019laguerre} to quantum information processing\cite{brandt2020high,hiekkamaki2021high,lib2022processing,lib2023resource}.    

In this work, we use a record-large programmable MPLC with 10 phase planes to sort multiple MUBs of spatially entangled photons for high-dimensional QKD. We demonstrate two high-dimensional QKD schemes using the same experimental setup. We first implement a five-dimensional QKD scheme with sorted measurements in all six MUBs, obtaining a secure key rate of 1.57 bits per sifted photon through semidefinite programming on the tomographically-complete data\cite{araujo2023quantum}. We then perform 25-dimensional QKD in two MUBs. We take advantage of the flexibility and programmability of the MPLC to circumvent the fact that arbitrary mode sorters in a 25-dimensional space are beyond state-of-the-art experimental capabilities. Instead of the standard computational and discrete Fourier transform (DFT) bases, we use the MPLC to measure the photons in tailored MUBs with a complexity that scales only with the square root of the dimension. We further show that the structure of such MUBs leads to a non-uniform experimental error distribution, which is beneficial to the secure key rate, measured to be 0.8 bits per sifted photon in the experiment. 

\section{Mode sorting with a multi-plane light converter}

Measurements in multiple MUBs are critical for QKD. Specifically, a mode sorter that can perform the required unitary transformations on the spatial modes that encode the state is crucial for increasing the secure key rate and must be realized for at least one MUB. This is in contrast to projective measurements with a success rate that is inversely proportional to the dimension.

\begin{figure}[h!]
\centering
\includegraphics[width=\columnwidth]{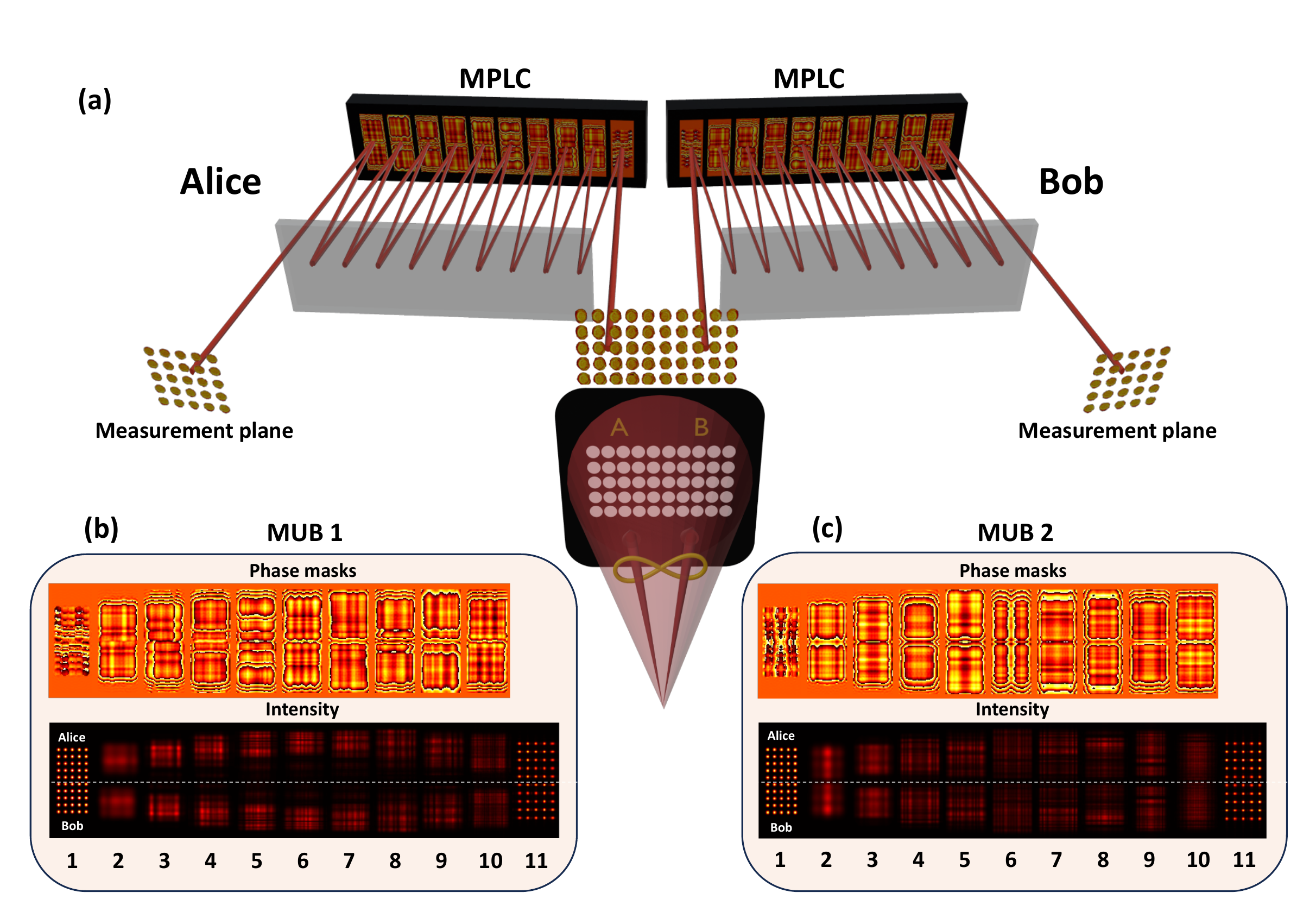}
     \caption{\textbf{Spatially-encoded QKD using multi-plane light conversion.} (a) Pairs of spatially entangled photons are generated via SPDC and are transmitted through a binary amplitude mask with 50 apertures. Alice uses a programmable MPLC with ten phase masks to measure one of the photons in one of several MUBs. Its twin photon is transmitted to Bob, who uses his programmable MPLC to sort the different spatial modes in one of the MUBs. Measurements in which Alice and Bob choose the same basis contribute to the establishment of a secret key. Examples of such measurements in two MUBs of dimension 25 are presented in panels (b) and (c), where the optimized MPLC phase masks and the sum of intensities from all input modes are plotted for Alice (top half) and Bob (bottom half) for all planes.
     }
 \label{fig1}
 \end{figure}

We realize a high-dimensional programmable mode sorter in multiple MUBs using a 10-plane MPLC (fig.\ref{fig1}a). For each MUB, the unitary transformation required for the change of basis is implemented by displaying different tailored phase masks on the MPLC planes. The phase masks are computed using the wavefront matching algorithm\cite{fontaine2019laguerre}. Examples of such phase masks used for 25-dimensional QKD are presented in fig.\ref{fig1}b,c together with the summed intensity from all input modes at each plane of the MPLC.

In our experiment, we use the MPLC for high-dimensional QKD (fig.\ref{fig1}a, see Supplementary). Alice produces pairs of spatially entangled photons via spontaneous parametric down-conversion (SPDC)\cite{walborn2010spatial}. The SPDC light passes at the far field through a binary amplitude mask with 50 apertures that define a 25-dimensional 'pixel-entangled' quantum state\cite{valencia2020high}. To encode the information, Alice chooses a MUB by programming the appropriate unitary transformation between the set of input and output spatial modes using the MPLC. She then measures her photon at one of the output modes, which determines the state of the photon sent to Bob. To decode the information, Bob uses a second MPLC to sort the modes in a chosen (transposed) MUB and measures his photon at the output. For experimental convenience, we realize the two independent MPLCs by bouncing the entangled photons ten times between a single SLM and a mirror, where the top part of each phase mask is used by Alice and the bottom part by Bob.

\section{Quantum key distribution in six mutually unbiased bases}

To demonstrate the flexibility of MPLCs, we first consider five spatial modes per photon and perform five-dimensional QKD with measurements in all six MUBs\cite{cerf2002security,mafu2013higher}. Along with the standard computational basis $\{\ket{n}\}_{n=1,\ldots,5}$ that can be measured without the MPLC, we choose five other MUBs $\left\{\ket{\psi^{(r)}_k}\right\}_{k=1,\ldots,5}$, with $k$ indexing the state and $r=2,\ldots,6$ the basis, such that\cite{wootters1989optimal}

\begin{equation} \label{mubs}
    \ket{\psi^{(r)}_k}=\frac{1}{\sqrt{5}}\sum^5_{n=1}e^{\frac{2\pi i}{5}[(k-1)(n-1)+(r-2)(n-1)^2)]}\ket{n}
\end{equation}

\begin{figure}[h!]
\centering
\includegraphics[width=\columnwidth]{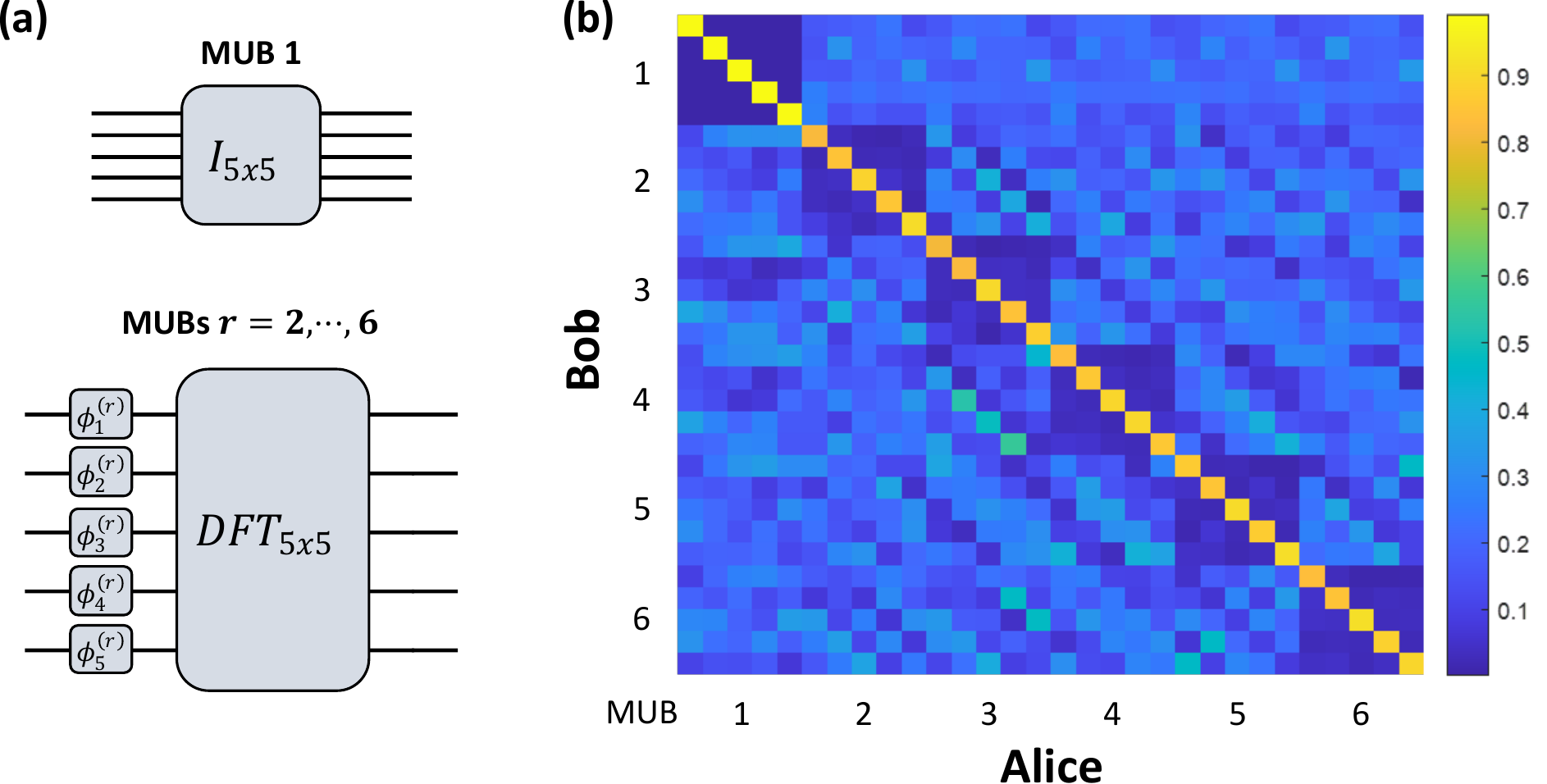}
     \caption{ \textbf{Five-dimensional QKD using a mode sorter in six MUBs.} We program the MPLC to sort the photons in all six MUBs of dimension five. The bases are chosen such that one basis is the trivial computational basis, while the others are measured by adding different relative phases at the first plane of the MPLC, before realizing an optical discrete Fourier transform (DFT)(a). We measure the correlations between the photons in all bases and estimate the probability of each outcome in Bob's measurement for each state sent by Alice (b). From these tomographically-complete measurements, we extract a secure key rate of 1.57 bits per sifted photon. 
     }
 \label{fig2}
 \end{figure}

For $r=2$, Eq. \ref{mubs} is exactly the Discrete Fourier Transform (DFT) basis of dimension five. Note that the four other bases $r=3,\ldots,6$ only differ from the DFT basis by phases that are independent of the state $k$. Therefore, bases $r=3,\ldots,6$ can be mapped onto the DFT basis by applying mode-dependent phases to the spatial modes. For our pixel-entangled states, this is easily implemented using a single phase mask at the first plane of the MPLC. Measurements in these MUBs are thus realized by encoding the DFT transformation using the MPLC and superimposing mode-dependent phases at the first plane of the MPLC to choose the desired MUB (fig.\ref{fig2}a).

The estimated probabilities for each of Bob's measurement outcomes given a state sent by Alice, in all MUBs, are presented in fig.\ref{fig2}b. From these probabilities, it is possible to compute a lower bound on the secure key rate. A rough lower bound can be obtained by applying the often-used simplifying assumption of identical depolarizing noise channels in all bases\cite{cerf2002security,sheridan2010security,bouchard2018experimental,djordjevic2019quantum}. This yields a secure key rate of $R=\log_2(d)-h_d(\frac{d+1}{d}E)-\frac{d+1}{d}E\log_2(d+1)=1.15 \pm 0.05$ bits per sifted photon, where $h_d(x)=-x\log_2(\frac{x}{d-1})-(1-x)\log_2(1-x)$ is the high-dimensional Shannon entropy and $E=11\%$ is the arithmetic mean of the error rates in all bases.

However, a tighter estimation of the lower bound could be obtained through numerical techniques\cite{coles2016numerical, winick2018reliable, sulimany2021fast, hu2022robust, araujo2023quantum} using the full tomographically-complete data in fig.\ref{fig2}b\cite{araujo2023quantum}. Indeed, using the method described in \cite{araujo2023quantum}, we observe a gradual increase in the estimated key rate when taking into account more of the experimental data, both when Alice and Bob measure in the same basis and when measuring in different bases (see Supplementary information). We obtain a secure key rate of 1.57 for the full tomographically-complete data, which is significantly higher than the loose bound given by assuming depolarizing noise.

\section{Experimentally-efficient 25-dimensional quantum key distribution}

Next, we scale up to implement 25-dimensional QKD in two MUBs. As 25-dimensional arbitrary mode sorters are still out of reach\cite{taballione2021universal}, experimental demonstrations at this scale can use a simple mode sorter to measure in the computational basis, and perform projective measurements in a second MUB\cite{mair2001entanglement}. The fact that the first MUB is the computational one forces the second MUB to involve a fully connected transformation that interferes all d modes (e.g. d-by-d DFT), which makes its realization challenging.

However, thanks to the flexibility of the MPLC to sort modes in different bases, we can design tailored measurement bases that reduce the experimental complexity. Recently, a new approach to reduce the experimental complexity of four-dimensional QKD has been proposed, where instead of the computational and DFT bases, one uses two MUBs in which only pairs of modes are interfered\cite{vagniluca2020efficient,da2021path}. We generalize this approach to any dimension $d=p^{2m}$, where $p$ is a prime number and $m>0$ is an integer. In this case, we find two MUBs that involve the interference of only $\sqrt{d}$ modes, significantly reducing the complexity of the measurement (see Supplementary information).

\begin{figure}[h!]
\centering
\includegraphics[width=\columnwidth]{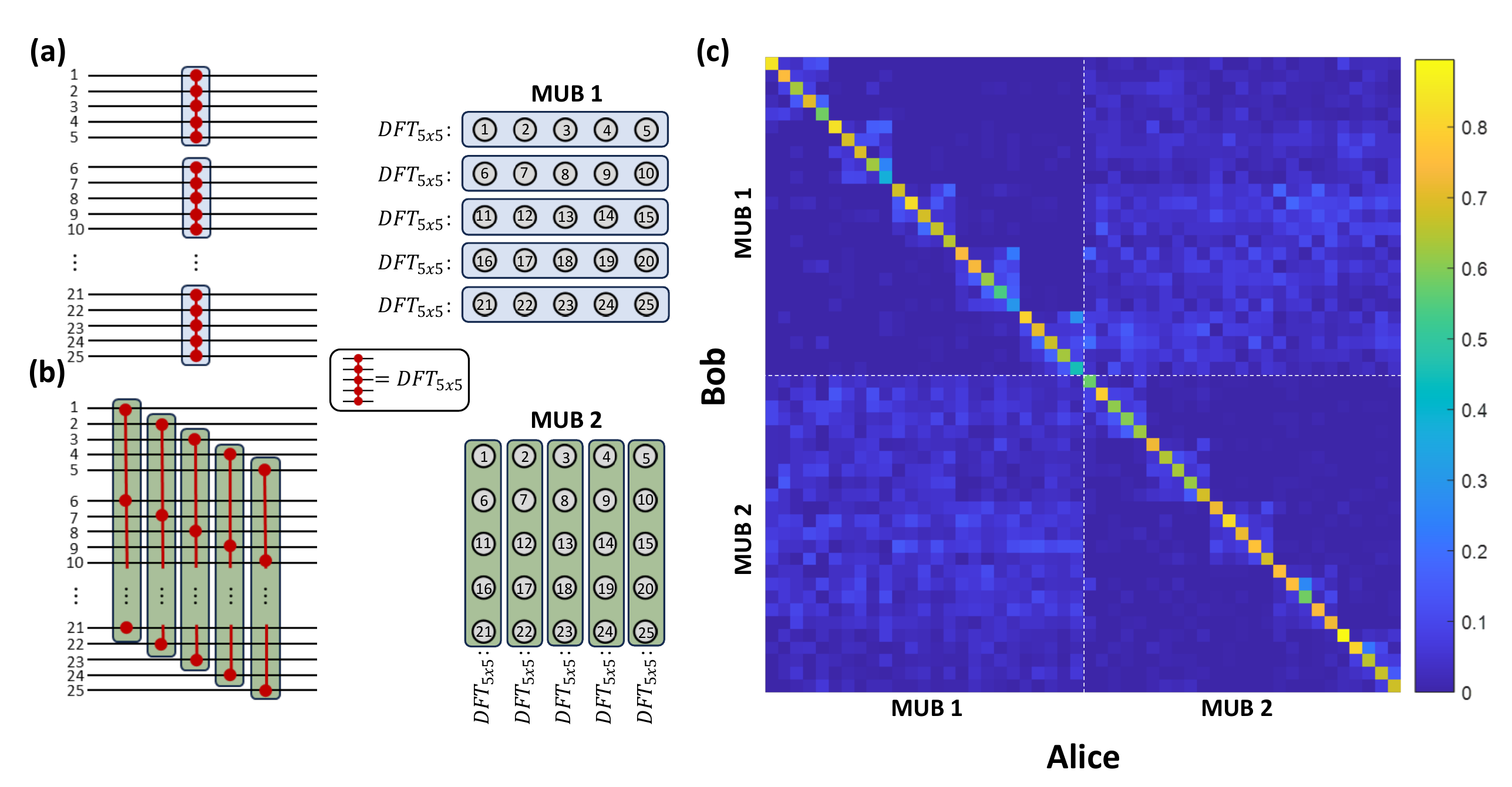}
     \caption{ \textbf{25-dimensional QKD in two MUBs.} We encode a 25-dimensional quantum state in a grid of 25 spatially non-overlapping modes. Instead of using the standard computational and 25-dimensional DFT bases, we design two MUBs that are tailored to the experimental capabilities and only require the interference of sets of five spatial modes (a),(b). Measurements in these MUBs are performed by applying five-dimensional DFT transformations along the rows (a) or columns (b) of the 5x5 grid of modes using the MPLC. The experimental results are presented in (c), yielding a secure key rate of 0.8 bits per photon.
     }
 \label{fig3}
 \end{figure}

We demonstrate this idea in dimension $d=5^2$ using our MPLC. The structure of the MUBs can be intuitively understood by placing the modes on a $\sqrt{d}$-by-$\sqrt{d}$ grid. Measurements in the two MUBs in this case are performed by applying $\sqrt{d}$-dimensional DFTs along the rows (MUB1, fig.\ref{fig3}a) or columns (MUB2, fig.\ref{fig3}b) of the grid. As any row and column only overlap at a single mode, these bases are mutually unbiased (see Supplementary information). We use the MPLC to experimentally realize such transformations on the 25 spatial modes of each photon and measure their correlations at the two MUBs (fig.\ref{fig3}c).

Interestingly, a non-uniform error distribution is clearly visible from the correlations in fig.\ref{fig3}c. Along with uniform error sources such as accidental coincidence counts, the structure of the unitary transformation, that interferes blocks of $\sqrt{d}$ modes, leads to a block-biased error distribution. Non-uniform error distributions are also common in other mode sets such as Lagurre Gaussian modes in free space \cite{krenn2014generation}, linearly polarized modes in multimode fibers \cite{ploschner2015seeing} or even time-bins \cite{sulimany2021fast}. In all cases, the block-biased noise behavior stems from the mode and MUB structures that dictate a lower probability of errors for distant modes that do not interfere. As the complexity of the semidefinite programming approach we used for $d=5$ prevents its implementation for $d=25$, in what follows, we utilize this block-biased noise structure to get a tighter analytical lower bound for the secure key rate.

In our experiment, the total error rate of $E_t=32.1\%$ is composed of a uniform error $E_u=7.3\%\approx0.23E_t$ and a 'block' error $E_b=24.8\%\approx0.77E_t$. The total uniform error $E_u$ was estimated by summing the probabilities for measuring a state in one of the $d-\sqrt{d}$ states outside the 'block' and multiplying by $\frac{d-1}{d-\sqrt{d}}$ to account for all $d-1$ error states. $E_b$ is then simply the remaining error probability within the $\sqrt{d}$-sized block. 

We take this into account in the secure key rate lower bound and obtain:

\begin{equation}
\begin{split} \label{Block biased}
    h_d(E_u,E_b)=-(1-E_t)\log_2(1-E_t)-\left(d-\sqrt{d}\right)\frac{E_u}{d-1}\log_2\left(\frac{E_u}{d-1}\right)\\
    -\left(\sqrt{d}-1\right)\left(\frac{E_b}{\sqrt{d}-1}+\frac{E_u}{d-1}\right)\log_2\left(\frac{E_b}{\sqrt{d}-1}+\frac{E_u}{d-1}\right)
\end{split}
\end{equation}

\begin{figure}[h!]
\centering
\includegraphics[width=\columnwidth]{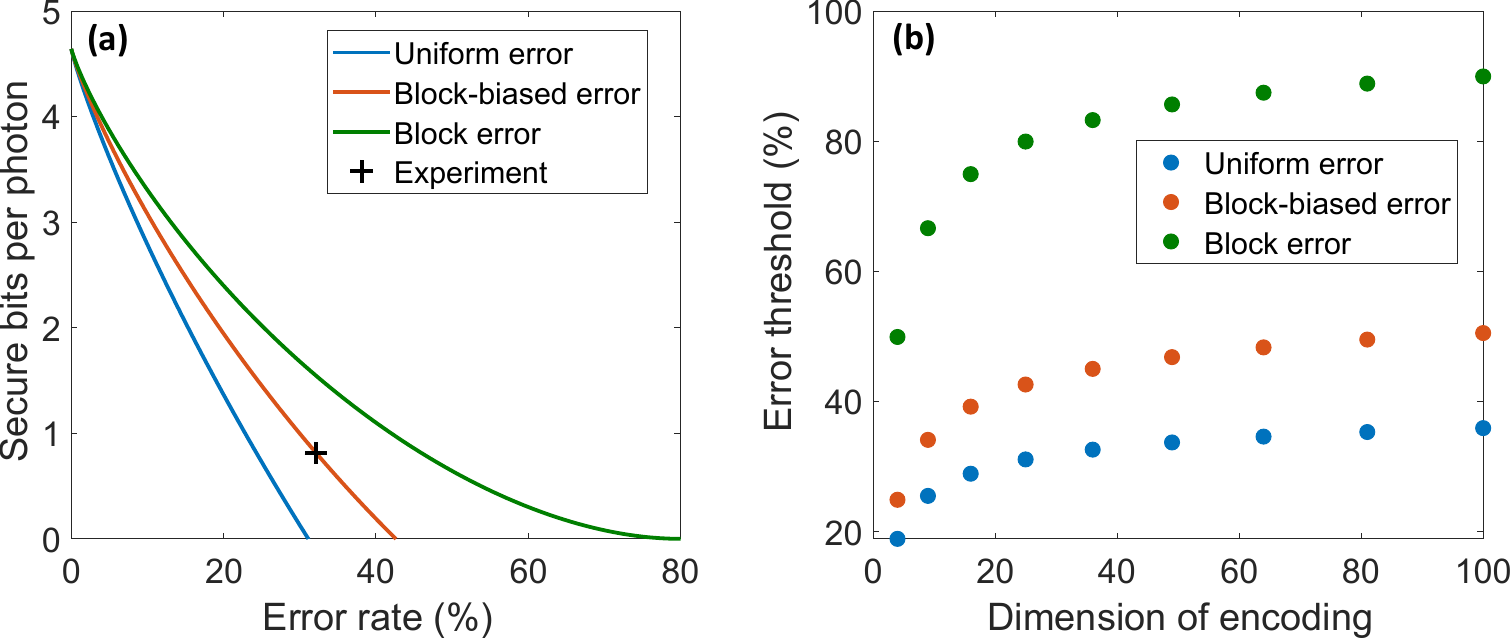}
     \caption{\textbf{Secret key rates and error threshold.} (a) The secret key rate per sifted photon in QKD is dependent on both the total error rate as well as the error distribution. In our 25-dimensional protocol, as a direct result of the block-like structure of the optical transformations, the errors are clearly not uniform and exhibit a strong block-like bias (orange curve). We therefore use a block-biased error model to calculate the secure key rate in this case, obtaining a rate of 0.8 bits per sifted photon (black marker). Non-uniform errors also increase the error threshold of the protocol (b). The error threshold increases significantly when the errors are block-biased, as in our experiment (orange), and even more so when they are completely within the block (green).
     }
 \label{fig4}
 \end{figure}

In Eq. \ref{Block biased}, the first term is for the measurement of the correct state, the second is for the $d-\sqrt{d}$ states outside the block, and the third is for the $\sqrt{d}-1$ possible error states within the block that are affected by both the uniform and block errors. Since uniform error maximizes the Shannon entropy, any non-uniformity in the error distribution increases the achievable secure key rate and total tolerable noise (fig.\ref{fig4}a,b). The secure key rate is lower-bounded by $R=\log_2(d)-2h_d(E_u,E_b)$\cite{berta2010uncertainty,sheridan2010security,djordjevic2019quantum}, yielding $0.8 \pm 0.07$ secure bits per sifted photon in our experiment. 

\section{Conclusion}

In this work, we have demonstrated high-dimensional QKD with spatially entangled photons by using MPLCs as programmable mode sorters. We have utilized the flexibility offered by our 10-plane MPLC to realize five-dimensional QKD with measurements in all six MUBs, and 25-dimensional QKD with measurements in two MUBs, in the same experimental setup.

For dimensions that are an even power of a prime number, we have further proposed pairs of MUBs for which the complexity of the experimental transformations only scales with the square root of the dimension. We have performed measurements in such MUBs for dimension 25 and shown that they lead to a non-uniform error distribution that allows for obtaining a tighter analytical lower bound for the secure key rate.

The main limitations of our proof-of-concept demonstration are the relatively high loss induced by the MPLC, which is currently in access of 10dB in our experiment (see Supplementary), and its slow refresh rate. These mainly result from the use of a programmable SLM and can be mitigated by using a constant phase mask instead\cite{fontaine2021hermite}. While the full programmability of our MPLC will be lost in this case, our choice of MUBs ensures that simple switching between different MUBs could still be achieved, at a lower loss and a faster rate. 

In the five-dimensional protocol, aside from the standard computational basis, switching between MUBs amounts to adding mode-dependent phases at a single plane, which can be realized using high-speed phase shifters. In the 25-dimensional protocol, the two MUBs only differ from one another by a rotation of the grid of modes by 90 degrees, which can be easily achieved using simple mirrors. 

We thus believe that incorporating our demonstration of high-dimensional QKD using MPLCs into state-of-the-art spatial high-dimensional QKD systems \cite{walborn2006quantum,etcheverry2013quantum,mirhosseini2015high,ding2017high,bouchard2017high,bouchard2018experimental,cozzolino2019orbital,tentrup2019large,zhou2019using,otte2020high,da2021path,hu2021pathways,ortega2021experimental,rozenman2023quantum,stasiuk2023high, forbes2024quantum} would pave the way towards high-rate long-range QKD.

\begin{backmatter}
\bmsection{Funding}
This research was funded by the Israeli Innovation Authority (Quantum Communication Consortium) and the  Israel Science Foundation (grants No. 2497/21 and 2137/19). O.L. acknowledges the support of the Clore Scholars Programme of the Clore Israel Foundation. K.S. acknowledges the support of the Israeli Council for Higher Education. Y.B. and K.S. acknowledge the support of the Zuckerman STEM Leadership Program. The research of M.A. was supported by the European Union--Next Generation UE/MICIU/Plan de Recuperación, Transformación y Resiliencia/Junta de Castilla y León.


\bmsection{Disclosures}

\noindent The authors declare no conflicts of interest.

\bmsection{Data availability} Data underlying the results presented in this paper are available in Ref. \cite{dataset}.

\bmsection{Supplemental document}
See Supplement 1 for supporting content.

\end{backmatter}


\bibliography{sample}

\def\thefigure{S\arabic{figure}}
\setcounter{figure}{0}
\renewcommand{\theequation}{S.\arabic{equation}}
\setcounter{equation}{0}
\renewcommand{\thetable}{S\arabic{table}}
\setcounter{table}{0}

\section{Supplementary information}
\subsection{Experimental setup}

\begin{figure}[htbp]
\centering
\includegraphics[width=\linewidth]{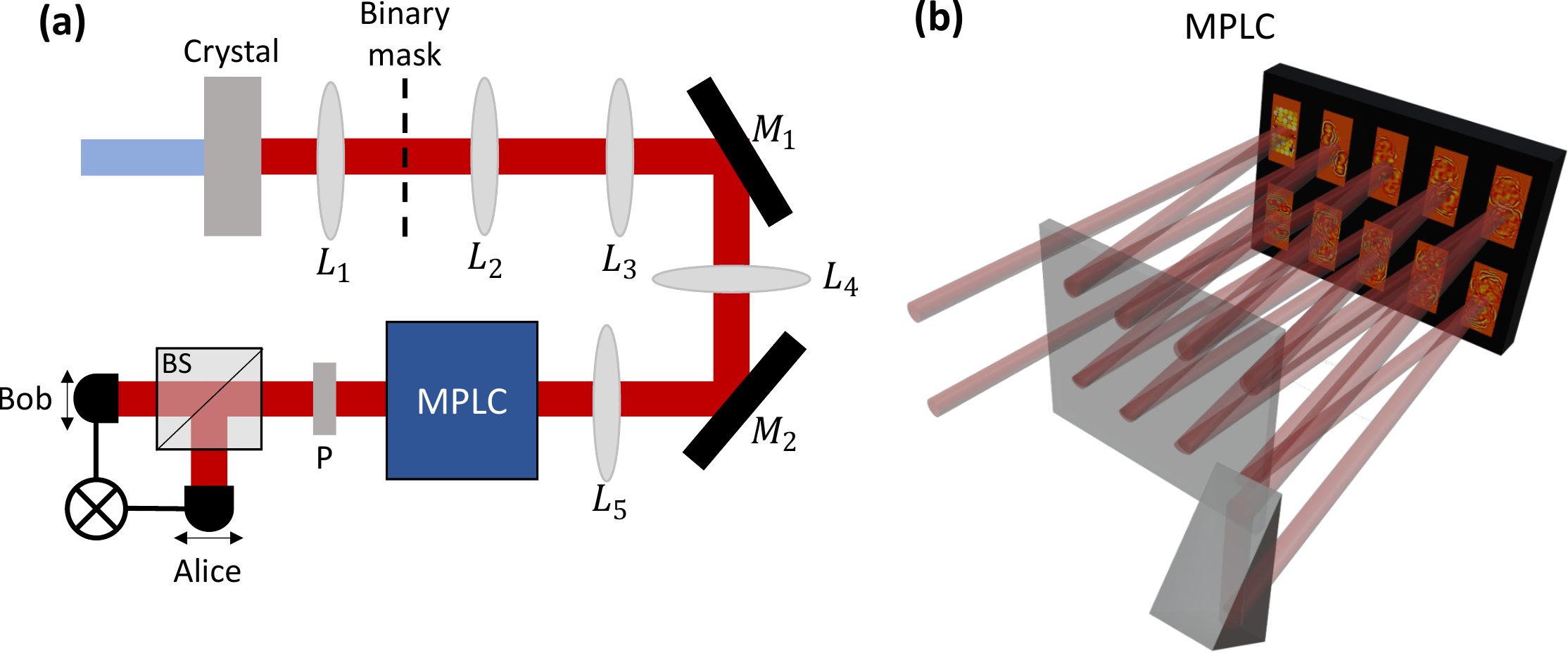}
\caption{(a) Illustration of the experimental setup. $L_1-L_5$- lenses, $M_1,M_2$- mirrors, BS- beam splitters, P- polarizer. A dichroic mirror that discards the pump beam after the crystal and 20nm-wide filters placed before the detectors are not shown. (b) The MPLC consists of an SLM, a mirror, and a right-angle prism. The entangled photons bounce ten times on the SLM. See text for more details.}
\label{fig:1}
\end{figure}

We generate spatially entangled photons via type-I spontaneous parametric down-conversion (SPDC) in a $8mm$-long Barium Borate (BBO) crystal. The pump beam is a $405nm$ continuous-wave laser (Cobolt 06-MLD), with a power of 125mW and a waist of $w\approx600\mu m$ at the crystal plane. For the 25-dimensional measurements, a lower power of 30mW is used to reduce the effect of accidental counts in our $400ps$ coincidence window. After a $f=150mm$ lens, a binary amplitude mask of 50 circular apertures with a radius of $100\mu m$ and a spacing of $300 \mu m$ defines the modes of our pixel-entangled state. From the binary amplitude mask, the entangled photons are then imaged onto the first plane of our 10-plane multi-plane light converter (MPLC, fig.\ref{fig:1}a). 

In each plane of the MPLC, half of the phase mask manipulates Alice's photon, and the other half independently manipulates Bob's photon. The two photons bounce between a spatial light modulator (SLM, Hamamatsu X13138-02) and a mirror ten times, bouncing off the top and bottom parts of the SLM five times each (fig.\ref{fig:1}b). The distance between the SLM and the mirror is $43.5 mm$ and the distance between the SLM and the right-angle prism used to redirect the light to the lower part of the SLM is $69 mm$. For each transformation, the ten 140-by-360 pixel phase masks are calculated using the wavefront matching algorithm with 30 iterations\cite{fontaine2019laguerre,lib2023resource}.

Two $100 \mu m$ fibers coupled to avalanche photo-diode single-photon detectors (Excelitas, SPCM-AQRH-62-FC) are scanned in the transverse plane to detect the correlations between the spatially entangled photons with an integration time of 100 seconds per measurement (Swabian instruments, Time Tagger 20). The fibers are located $43.5 mm$ after the last plane of the MPLC, effectively at plane '11'. For each mutually unbiased basis (MUB), the probabilities for Bob's different measurement outcomes given a state sent by Alice are obtained from the raw correlations by normalizing them by the sum of coincidence counts for the given sent state.

The total loss of the MPLC is dependent on the applied transformation and is estimated by measuring the coincidence rate before and after the MPLC. We observe average losses of 10.7dB and 13.4dB per photon for the five-dimensional and 25-dimensional measurements, respectively, which could be significantly reduced in the future by using non-programmable MPLCs.

\subsection{Mutually unbiased bases with square-root complexity}

Measuring a $d$-dimensional photon in a given MUB requires the realization of an appropriate change-of-basis unitary transformation. The complexity of such measurements thus grows with the number of interfering modes $m$, as the number of planes in the MPLC or components in a photonic integrated circuit grows linearly with $m$. This means that the total number of components (e.g. phase shifters) scales with $md$, and the transmission of the device, which is determined by the number of planes or the depth of the photonic circuit, drops as $t^m$ where $t$ is the effective transmission of each plane or layer.

Typically, the standard computational basis and the $d$-dimensional Discrete Fourier Transform (DFT) basis are used in high-dimensional quantum key distribution (QKD). While measurements in the computational basis are often trivial, realizing the $d$-dimensional DFT requires a $d$-by-$d$ unitary transformation that interferes $m=d$ modes.

\begin{figure}[htbp]
\centering
\includegraphics[width=\linewidth]{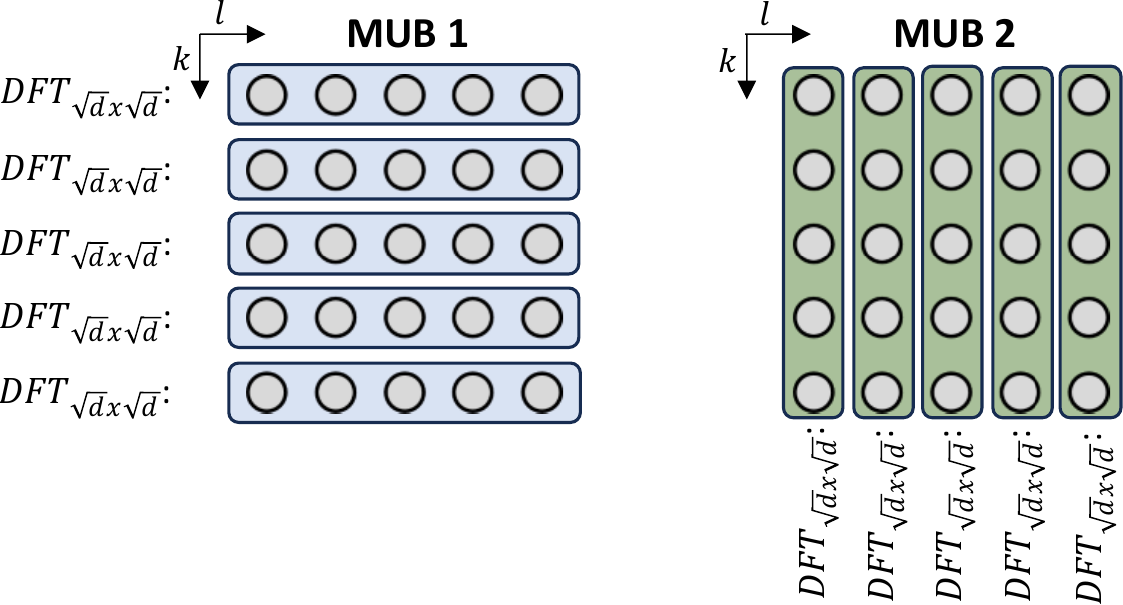}
\caption{Illustration of the two MUBs at dimension $d$. The bases are obtained from the computational basis encoded on a grid by applying DFTs along the rows or columns.}
\label{fig:2}
\end{figure}

For a dimension $d=p^{2m}$, where $p$ is a prime number and $m>0$ is an integer, we discuss in the main text the design of a pair of MUBs, both consisting of sets of only $m=\sqrt{d}$ interfering modes. Here, we elaborate on the construction of such bases and write their states explicitly. For that, it is convenient to order the $d$ states in the computational basis on a $\sqrt{d}$-by-$\sqrt{d}$ grid. Each state is labeled with its coordinates on the grid, $(k,l)$, so that the computational basis is written as $\left\{\ket{k,l}\right\}_{k=1,\ldots,\sqrt{d},l=1,\ldots,\sqrt{d}}$ (fig.\ref{fig:2}).

Now, two MUBs, $\left\{\ket{\psi^{(1/2)}_{a,b}}\right\}_{a=1,\ldots,\sqrt{d},b=1,\ldots,\sqrt{d}}$, can be constructed by taking the $\sqrt{d}$-DFTs along the rows (MUB1) or columns (MUB2) of the grid. This yields two MUBs of the form:
\begin{gather}
\ket{\psi^{(1)}_{a,b}}=\frac{1}{d^{0.25}}\sum^{\sqrt{d}}_{l=1} e^{\frac{2\pi i}{\sqrt{d}}(b-1)(l-1)}\ket{a,l} \label{eq:MUB1} \\ 
\ket{\psi^{(2)}_{a,b}}=\frac{1}{d^{0.25}}\sum^{\sqrt{d}}_{k=1} e^{\frac{2\pi i}{\sqrt{d}}(a-1)(k-1)}\ket{k,b} \label{eq:MUB2}
\end{gather}
which only involve the superposition of $\sqrt{d}$ modes in each state, simplifying the experimental realization considerably.

\subsection{Computation of the key rate using full data}

In order to compute the key rate using the full tomographically-complete data, we use the technique proposed in \cite{araujo2023quantum}. The first step in the technique is to infer a confidence region in the space of quantum states from the measurement data, and the second is to minimize the key rate over this region using an SDP (semidefinite programming) hierarchy.

The first step is necessary for two reasons: first, we do not want an estimate of the key rate but rather a lower bound. Second, a naïve linear inversion of the data does not result in a valid quantum state, which not only is a bad estimate but also breaks a fundamental assumption of QKD. To obtain the confidence region, we use Bayesian parameter estimation with a flat prior and a standard confidence level of 0.95.

Using this confidence region, we then minimize the key rate using five different subsets of the data: the average error when Alice and Bob are measuring in the same basis, all the errors when measuring in the same basis, all the errors when measuring in all bases, all probabilities when measuring in the same bases, and the full data. For clarity, these subsets are specified in the following equations
\begin{subequations}\label{eq:subsets}
\begin{gather}
E = 1 - \frac{1}{d+1}\sum_{k=0}^d\sum_{a=0}^{d-1} p(a,a|k,k) \\
E_k = 1 - \sum_{a=0}^{d-1} p(a,a|k,k) \quad \forall k\\
E_{k,l} = 1 - \sum_{a=0}^{d-1} p(a,a|k,l) \quad \forall k,l\\
E^{a,b}_{k}  = p(a,b|k,k) \quad \forall a,b,k\\
E^{a,b}_{k,l}  = p(a,b|k,l)\quad \forall a,b,k,l
\end{gather}
\end{subequations}
Here $p(a,b|k,l)$ is the probability of Alice and Bob obtaining results $a,b$ when measuring in MUBs $k,l$ (with Bob's MUBs being transposed). The results are showing in Table \ref{tab:key_rates}. We can see that, as expected, the more data we take into account the higher secure key rate we can certify.
\begin{table}
    \centering
    \begin{tabular}{cc}
        Key rate & subset \\
        1.3881 & $E$ \\
        1.4048 & $E_k$ \\
        1.4879 & $E_{k,l}$\\
        1.5219 & $E^{a,b}_{k}$\\
        1.5733 & $E^{a,b}_{k,l}$
    \end{tabular}
    \caption{Key rates for the subsets of the data specified in equation \eqref{eq:subsets}}
    \label{tab:key_rates}
\end{table}

\end{document}